\documentclass[%
superscriptaddress,
showpacs,preprintnumbers,
amsmath,amssymb,
aps,
prl,
twocolumn
]{revtex4-1}
\usepackage{ulem}
\usepackage{graphicx}
\usepackage{dcolumn}
\usepackage{bm}
\usepackage{color}

\begin {document}

\title{Heterogeneous biological membranes regulate protein partitioning via fluctuating diffusivity}

\author{Ken Sakamoto}
\affiliation{%
Department of System Design Engineering, Keio University, Yokohama, Kanagawa 223-8522, Japan
}%

\author{Takuma Akimoto}
\affiliation{%
Department of Physics, Tokyo University of Science, Noda, Chiba 278-8510, Japan
}%

\author{Mayu Muramatsu}
\affiliation{%
Department of Mechanical Engineering, Keio University, Yokohama, Kanagawa 223-8522, Japan
}%

\author{Mark S. P. Sansom}
\affiliation{%
Department of Biochemistry, University of Oxford, South Parks Road, Oxford OX1 3QU, U.K.
}%

\author{Ralf Metzler}
\affiliation{%
Institute of Physics \& Astronomy, University of Potsdam, 14476 Potsdam-Golm, Germany
}%
\affiliation{%
Asia Pacific Centre for Theoretical Physics, Pohang, 37673, Republic of Korea
}

\author{Eiji Yamamoto}
\email{eiji.yamamoto@sd.keio.ac.jp}
\affiliation{%
Department of System Design Engineering, Keio University, Yokohama, Kanagawa 223-8522, Japan
}%



\begin{abstract}
Cell membranes phase separate into ordered ${\rm L_o}$ and disordered ${\rm L_d}$ domains depending on their compositions.
This membrane compartmentalization is heterogeneous and regulates the localization of specific proteins related to cell signaling and trafficking.
However, it is unclear how the heterogeneity of the membranes affects the diffusion and localization of proteins in ${\rm L_o}$ and ${\rm L_d}$ domains.
Here, using Langevin dynamics simulations coupled with the phase-field (LDPF) method, we investigate several tens of milliseconds-scale diffusion and localization of proteins in heterogeneous biological membrane models showing phase separation into ${\rm L_o}$ and ${\rm L_d}$ domains.
The diffusivity of proteins exhibits temporal fluctuations depending on the field composition.
Increases in molecular concentrations and domain preference of the molecule induce subdiffusive behavior due to molecular collisions by crowding and confinement effects, respectively.
Moreover, we quantitatively demonstrate that the protein partitioning into the ${\rm L_o}$ domain is determined by the difference in molecular diffusivity between domains, molecular preference of domain, and molecular concentration.
These results pave the way for understanding how biological reactions caused by molecular partitioning may be controlled in heterogeneous media.
Moreover, the methodology proposed here is applicable not only to biological membrane systems but also to the study of diffusion and localization phenomena of molecules in various heterogeneous systems.
\end{abstract}

\maketitle

Biological membranes are composed of various kinds of proteins and lipids.
Differences in the molecular composition relate to rich patterns of phase separation~\cite{FanSammalkorpiHaataja2010a, LeventalGrzybekSimons2011, SanchezTricerriGratton2012, SezginGutmannBuhlDirkxGrzybekCoskunSolimenaSimonsLeventalSchwille2015, SezginLeventalMayorEggeling2017}.
Mixtures of saturated and unsaturated lipids generally cause phase separation into liquid-ordered (${\rm L_o}$) and liquid-disordered (${\rm L_d}$) domains~ \cite{Pike2006, HeberleWuGohPetruzieloFeigenson2010, DeWitDanialKukuraWallace2015}.
Specifically, the ${\rm L_d}$ domain is rich in unsaturated lipids and of high fluidity, while the ${\rm L_o}$ domain is rich in saturated lipids and of low fluidity.
${\rm L_o}$ domains, enriched in sphingolipids and cholesterol, are often referred to as lipid rafts~\cite{LingwoodSimons2010}, and are thought to play a crucial role in a variety of cellular processes such as cell signaling and trafficking.
Lipid rafts are generally considered to be small, heterogeneous, and highly dynamic domains of several tens of nanometers size with estimated life time $0.1$--$10^2$~s~\cite{PralleKellerFlorinSimonsHoerber2000, KomuraSuzukiAndoKonishiKoikedaImamuraChaddaFujiwaraTsuboiShengChoFurukawaFurukawaYamauchiIshidaKusumiKiso2016, WuLinYenHsieh2016, KusumiFujiwaraTsunoyamaKasaiLiuHirosawaKinoshitaMatsumoriKomuraAndoSuzuki2020}.
The coexistence of ${\rm L_o}$ and ${\rm L_d}$ domains has been observed in synthetic model membranes under external stimuli or specific thermodynamic conditions.
Direct imaging of cell-derived plasma membranes of giant plasma membrane vesicles (GPMVs) has also confirmed the presence of nanodomains~\cite{GeGidwaniBrownHolowkaBairdFreed2003, LingwoodRiesSchwilleSimons2008, VeatchCicutaSenguptaHonerkamp-SmithHolowkaBaird2008, LeventalByfieldChowdhuryGaiBaumgartJanmey2009, KaiserLingwoodLeventalSampaioKalvodovaRajendranSimons2009, LiWangKakudaLondon2020, HeberleDoktorovaScottSkinkleWaxhamLevental2020} at or near physiological temperature.
Although there has been a longstanding debate regarding the nature and biological role(s) of these domains in living cells~\cite{LeventalLeventalHeberle2020, ShawGhoshVeatch2021}, a large number of recent studies have provided evidence for the coexistence of these domains in intact cells~\cite{ToulmayPrinz2013, StoneShelbyNunezWisserVeatch2017, KinoshitaSuzukiMatsumoriTakadaAnoMorigakiAbeMakinoKobayashiHirosawaFujiwaraKusumiMurata2017, Koyama-HondaFujiwaraKasaiSuzukiKajikawaTsuboiTsunoyamaKusumi2020, UrbancicSchiffelersJenkinsGongSantosSchneiderOBrienBallVuongAshmanSezginEggeling2021, BagWagenknechtWiesnerLeeShiHolowkaBaird2021, ShelbyCastelloSerranoWisserLeventalVeatch2023}.
These studies investigated the recruitment and exclusion of various probes associated with clustered proteins within cell membranes, and demonstrated that the concentration of probes in clusters reflects the partitioning observed in phase-separated domains.
The ${\rm L_o}$ domains are formed not only by lipids but also by protein--lipid complexes, where the detailed properties, such as size, lifetime, and stability, depend on their composition and interaction with scaffolding proteins~\cite{LeventalLeventalHeberle2020, ShawGhoshVeatch2021}.

In terms of lateral diffusion of membrane proteins, phase separation may be considered as presenting an inhomogeneous field in which the protein molecules diffuse.
The diffusivities of molecules in such inhomogeneous fields are known to be non-uniform in time and space~\cite{Torreno-PinaManzoGarcia-Parajo2016, MetzlerJeonCherstvy2016}.
Experimental techniques, such as stimulated emission depletion microscopy combined with fluorescence correlation spectroscopy (STED-FCS) and single-particle tracking (SPT), have revealed dynamically heterogeneous motion of proteins in biological membranes~\cite{SergeBertauxRigneaultMarguet2008, WeigelSimonTamkunKrapf2011, ManzoTorreno-PinaMassignanLapeyreLewensteinGarcia2015, WuLinYenHsieh2016, HeSongSuGengAckersonPengTong2016, WeronBurneckiAkinSoleBalcerekTamkunKrapf2017, SadeghHigginsMannionTamkunKrapf2017, CharalambousMunoz-GilCeliGarcia-ParajoLewensteinManzoGarcia-March2017, SilMateosNathBuschowManzoSuzukiFujiwaraKusumiGarcia-ParajoMayor2020, ChaiChengLiaoLinHsieh2022}.
Particularly, the local diffusivity of tracers fluctuates significantly with time due to the influence of heterogeneity in the field, e.g. intermittent trapping in domains~\cite{SergeBertauxRigneaultMarguet2008, WuLinYenHsieh2016}, transient interactions with partners~\cite{ManzoTorreno-PinaMassignanLapeyreLewensteinGarcia2015, CharalambousMunoz-GilCeliGarcia-ParajoLewensteinManzoGarcia-March2017}, or slow-active remodeling of the underlying cortical actin network~\cite{HeSongSuGengAckersonPengTong2016, SadeghHigginsMannionTamkunKrapf2017}.
However, due to the difficulty of simultaneous measurement of molecular motion and field heterogeneity, the precise effects of membrane heterogeneity on molecular diffusion and distribution remain obscure.
Although many theoretical models on molecular diffusion with fluctuating diffusivity have been proposed to explain the characteristics of non-Gaussian behavior and anomalous diffusion
\cite{MassignanManzoTorreno-PinaGarcia-ParajoLewensteinLapeyre2014, ChubynskySlater2014, UneyamaMiyaguchiAkimoto2015, AkimotoYamamoto2016, MiyaguchiAkimotoYamamoto2016, CherstvyMetzler2016, ChechkinSenoMetzlerSokolov2017, TyagiCherayil2017, JainSebastian2018, SabriXuKrapfWeiss2020, Hidalgo-SoriaBarkai2020, SposiniGrebenkovMetzlerOshaninSeno2020, BarkaiBurov2020, WangSenoSokolovChechkinMetzler2020, Pacheco-PozoSokolov2021}, it is important to understand the origin of fluctuations at the molecular level, specifically how phase separation, modeled as an inhomogeneous field, affects protein diffusivity and promotes protein crowding, or how molecular crowding induces phase separation and expands nanoscale domains.
This understanding will help to clarify the role of protein--lipid and protein--protein interactions in the signaling process.

Molecular dynamics (MD) simulations have provided molecular details on protein diffusion in biological systems~\cite{GooseSansom2013, PartonTekBaadenSansom2013, JavanainenHammarenMonticelliJeonMiettinenMartinez-SearaMetzlerVattulainen2013, JavanainenMartinez-SearaMetzlerVattulainen2017, JeonJavanainenMartinez-SearaMetzlerVattulainen2016, YamamotoAkimotoKalliYasuokaSansom2017, YamamotoAkimotoMitsutakeMetzler2021}, and revealed temporal fluctuating of the protein diffusivity due to protein--protein and protein--lipid interaction~\cite{JeonJavanainenMartinez-SearaMetzlerVattulainen2016, YamamotoAkimotoKalliYasuokaSansom2017}.
However, it remains a challenge for simulations to directly inform molecular dynamics on a spatiotemporal scale comparable to experiments.
Here, using a mesoscale simulation technique, we unveil diffusion properties and distributions of molecules in heterogeneous biological membrane models.
This coarse-grained level, combining Langevin dynamics simulations and phase-field (LDPF) methods, capture the motion of individual molecules in heterogeneous membranes at several tens of milliseconds timescales.
We show the existence of fluctuating diffusivity and a distribution of molecules in heterogeneous membranes depending on various parameters such as heterogeneity of fields, molecular concentrations, and domain preference of molecules.
This coarse-grained approach allows us to disentangle the effects of individual parameters on the observed protein motion, e.g. the diffusivity difference between the two membrane phases, the area density covered by proteins, or the protein affinity to a specific membrane domain.
These results will be important to inform future experiments in real membrane systems in which some effects may be obscured by the complexity of the system.

\begin{figure*}[tb]
\centering
\includegraphics[width = 150 mm,bb= 0 0 971 649]{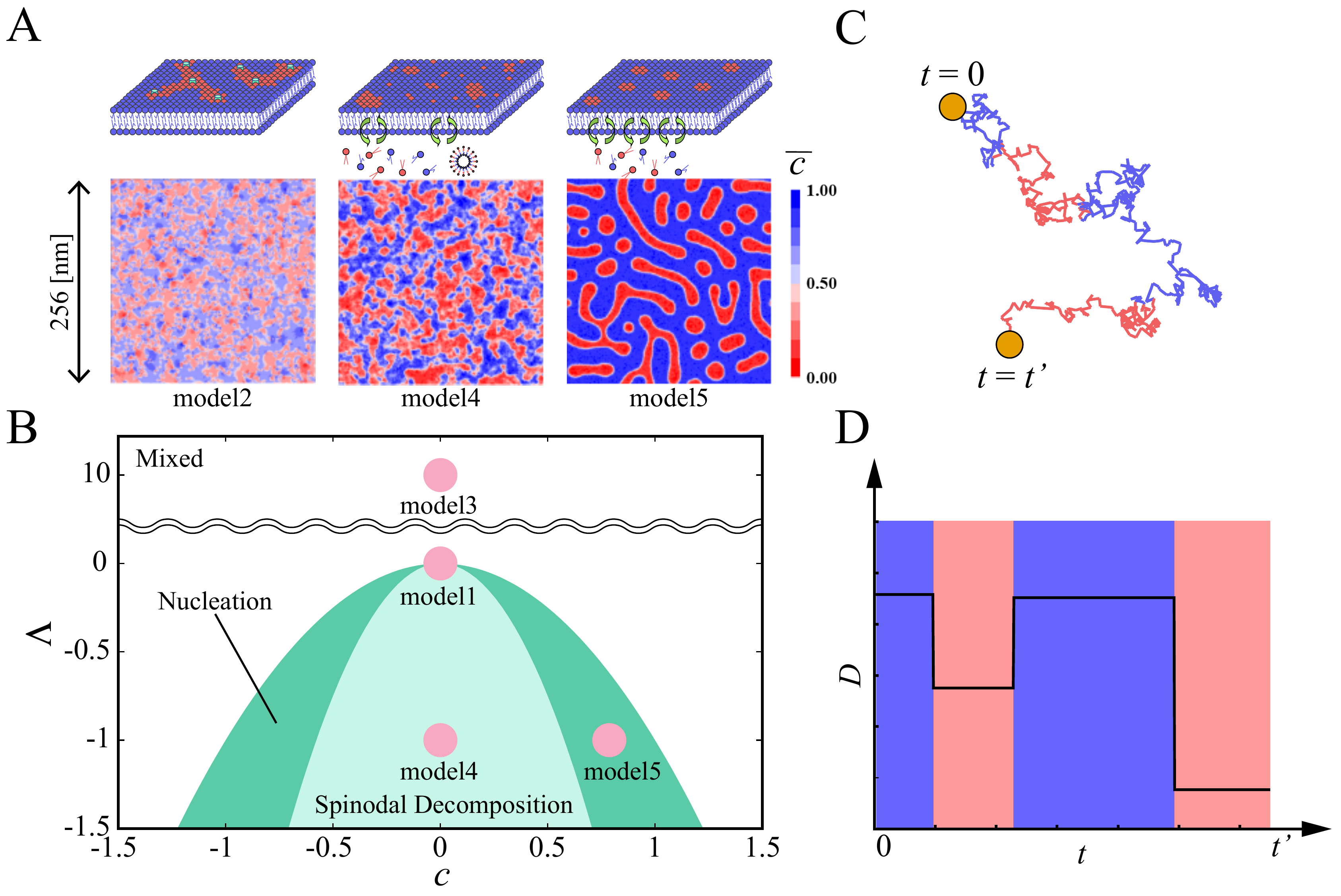}
\caption{Diffusivity fluctuations in heterogeneous biological membrane models.
(A)~Snapshots of normalized $c$ field configuration, $0 < \overline{c} < 1$, from phase-field simulations of heterogeneous biological membrane models~\cite{FanSammalkorpiHaataja2010, FanSammalkorpiHaataja2010a}, (model2)~interface pinning by proteins, (model4)~recycling in immiscible system, and (model5)~coupling to lipid reservoir.
Red and blue colored regions represent ${\rm L_o}$ and ${\rm L_d}$ domains, respectively.
(B)~Phase diagram for the models; the temperature difference $\Lambda$ from the critical temperature at which phase separation occurs v.s. $c$.
Considering a free energy term $F$ with $\alpha = 0$ in eq.~\ref{eq:free_energy}, the phase separation is classified as ``Mixed'', ``Nucleation'', and ``Spinodal Decomposition''.
(C)~Trajectory and (D)~the corresponding fluctuating diffusivity of a molecule depending on the $\overline{c}$ field.
The red and blue colors represent slow and fast diffusive states in ${\rm L_o}$ and ${\rm L_d}$ domains, respectively.
Averaged diffusion coefficients are shown for each state.}
\label{fig1}
\end{figure*}

\begin{figure*}[tb]
\centering
\includegraphics[width = 140 mm,bb= 0 0 898 761]{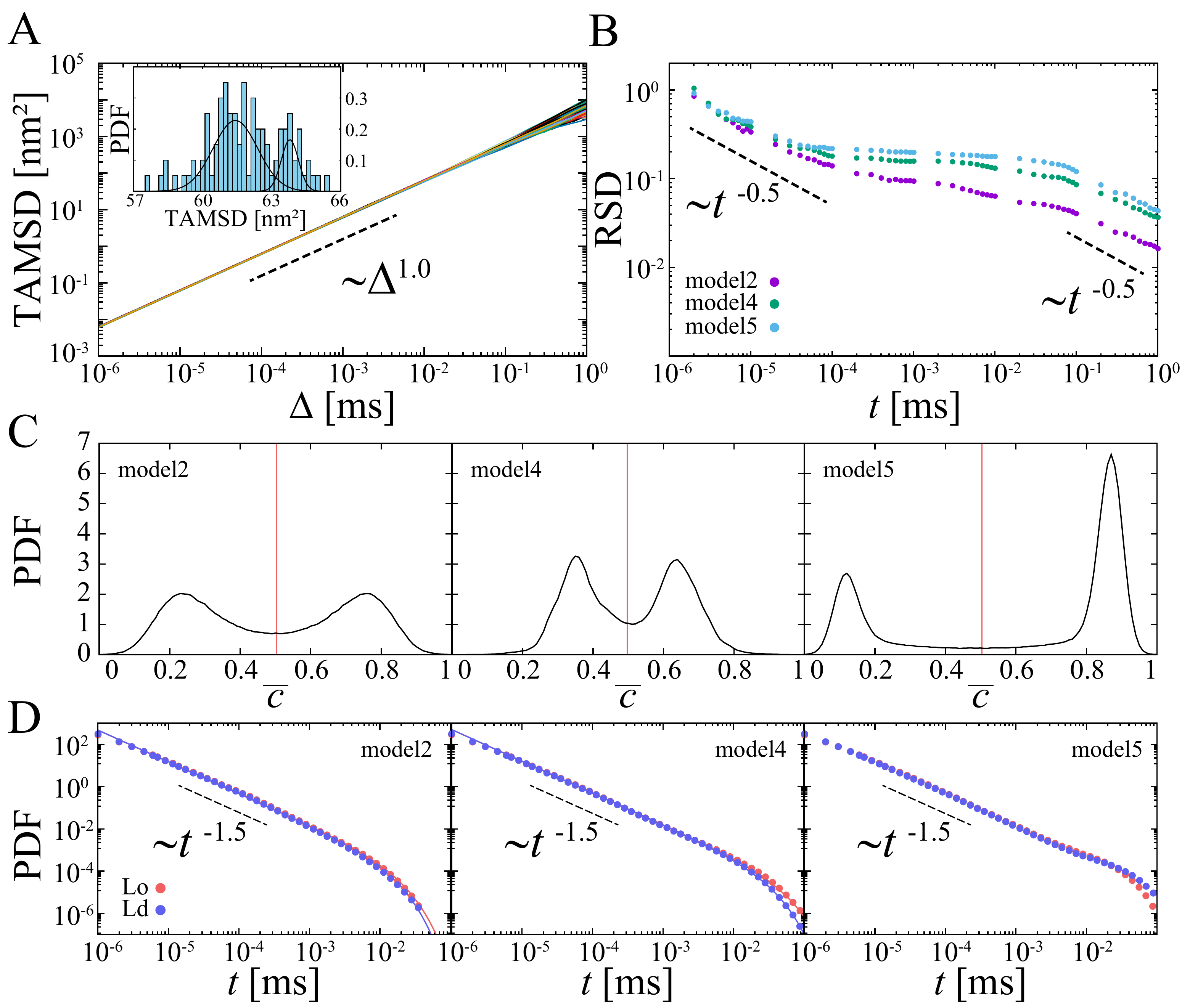}
\caption{Diffusion of an isolated protein molecule in heterogeneous biological membrane models.
(A)~TAMSDs of 100 trajectories of a molecule duffuising in the Model5 membrane for measurement time $t = 10$~ms.
The histogram shows the distribution of TAMSDs at $\Delta = 10^{-2}$~ms.
(B)~RSDs of TAMSDs for three different membrane models with $\Delta = 10^{-6}$~ms.
The RSD was calculated from 100 trajectories for each model.
(C)~Distribution of normalized order parameter $\overline{c}$.
The red colored line is the boundary defining ${\rm L_o}$ ($\overline{c}<0.5$) and ${\rm L_d}$ ($\overline{c} \geq 0.5$) domains.
(D)~Distribution of residence time of the particle in ${\rm L_o}$ and ${\rm L_d}$ domains following the power-law trend $t^{-1.5}$ with a log time cutoff at around $10^{-2}$~ms.
Dashed lines are shown as a reference for power-law decay.}
\label{fig2}
\end{figure*}

\section*{Results}
\subsection*{Fluctuating diffusivity of an isolated molecule in heterogeneous biological membrane models}
In our analysis, we focused on three distinct phase-separated heterogeneous biological membrane models described in previous studies~\cite{FanSammalkorpiHaataja2010}.
The phase separation process is measured in terms of the field $c({\bm r},t)$, the deviation of the local composition from the critical composition (see Methods for the simulation details).
The ordered ($c<0$) and disordered ($c>0$) phases denote the raft (${\rm L_o}$) and non-raft (${\rm L_d}$) domains, respectively.
The distribution $c({\bm r},t)$ can be obtained by solving the reaction-diffusion equation.
The specific model choices for the parameters induce clear phase separation and represent lipid raft formation; (model2) interface pinning by immobile membrane proteins~\cite{YethirajWeisshaar2007, LaradjiGuoGrantZuckermann1992}, (model4) immiscible lipid systems, and (model5) coupling to lipid reservoir~\cite{GomezSaguesReigada2008,Foret2005, FanSammalkorpiHaataja2010a} (see Fig.~\ref{fig1}A).
Considering the free energy term $F$, the phase separation is classified as ``Mixed'', ``Nucleation'', and ``Spinodal Decomposition''  (see Fig.~\ref{fig1}B).
Since the ${\rm L_o}$ and ${\rm L_d}$ domains have different compositions, diffusion coefficients of the biomolecules are different~\cite{WuLinYenHsieh2016, KusumiFujiwaraTsunoyamaKasaiLiuHirosawaKinoshitaMatsumoriKomuraAndoSuzuki2020}.
To describe the diffusion of target protein molecules in such heterogeneous media, we considered the Langevin equation with fluctuating diffusivity,
\begin{equation}
    \frac{d{\bm r}(t)}{dt} = \sqrt{2D({\bm r}(t),t)}{\bm w(t)},
    \label{eq:langevin}
\end{equation}
where ${\bm r}(t)$ is the position of a diffusing molecule at time $t$, and ${\bm w(t)}$ is white Gaussian noise with $\langle {\bm w(t)} \rangle = 0$.
The diffusion coefficient $D({\bm r}(t),t)$ varies depending on the field composition, $D({\bm r}(t),t) = (c_b + \overline{c({\bm r}(t),t)}) D_0$, where $\overline{c({\bm r}(t),t)}$ is the normalized order parameter field ($0 < \overline{c} <1$) (see Figs.~\ref{fig1}CD for a sample trajectory).
For a single molecular system, $c_b = 1$ and $D_0 = 1$ were used in each model, i.e. $D({\bm r}(t),t)$ fluctuates in the range of 1 to 2.
The simulation time step $dt = 0.001$ and $D_0 = 1$ correspond to the physical quantities of 1~ns and 1~$\mu$m$^2$/s, respectively. 
In the simulations, the system size $L$ corresponds to 256~nm with periodic boundary conditions.
Simulations were performed for $10^7$ steps corresponding to $10$~ms and analyzed after $10^6$ steps (1~ms) of reaching equilibrium.
Because the lifetime of the raft domain is $0.1$--$10^2$~s~\cite{PralleKellerFlorinSimonsHoerber2000, KomuraSuzukiAndoKonishiKoikedaImamuraChaddaFujiwaraTsuboiShengChoFurukawaFurukawaYamauchiIshidaKusumiKiso2016, WuLinYenHsieh2016, KusumiFujiwaraTsunoyamaKasaiLiuHirosawaKinoshitaMatsumoriKomuraAndoSuzuki2020}, we here fixed the field variation and focused on time scales shorter than the field variation.
This can allow us to evaluate the effect of spatial heterogeneity on the molecular dynamics, specifically elucidating how the (pre-existing) raft domains affects the behavior of other molecules.

First, we calculated the time-averaged mean squared displacement (TAMSD) as a quantity that characterizes the global diffusivity (see Fig.~\ref{fig2}A),
\begin{equation}
    \overline{\delta {\bm r}^2(\Delta;t)} = \frac{1}{t - \Delta}
    \int^{t - \Delta}_0 [{\bm r}(t' + \Delta) - {\bm r}(t')]^2 dt',
\end{equation}
where $\Delta$ is a lag time and $t$ is a measurement time.
Individual TAMSDs increase linearly and show some amplitude scatter.
The probability density function (PDF) of TAMSDs at $\Delta = 10^{-2}$ ms is found to have a distribution with roughly two peaks.
This scatter is considered to be an effect of the inhomogeneity of the concentration distribution in the field.

In order to quantitatively evaluate the effect of different patterns of heterogeneity on the diffusivity fluctuations, the relative standard deviation (RSD) of the TAMSDs was analyzed,
\begin{equation}
    {\rm RSD} = \frac{\sqrt{\langle \overline{\delta^2(\Delta;t)}^2 \rangle
            - \langle \overline{\delta^2(\Delta;t)} \rangle ^2}}
    {\langle \overline{\delta^2(\Delta;t)} \rangle}.
\end{equation}
It is known that RSD decays as $t^{-0.5}$ in ergodic diffusion, e.g. Brownian motion.
In the case of non-ergodic diffusion processes, e.g. the continuous-time random walk~\cite{HeBurovMetzlerBarkai2008, MiyaguchiAkimoto2011}, the RSD converges to a nonzero value for all $\Delta \ll t$ as $t \rightarrow \infty$.  
In fluctuating diffusivity models~\cite{UneyamaMiyaguchiAkimoto2015, MiyaguchiAkimotoYamamoto2016, AkimotoYamamoto2016, AkimotoYamamoto2016a}, the RSD exhibits a crossover from a plateau to a $t^{-0.5}$ decay with a long crossover time.
Here, the RSD shows a plateau in the time region $t \sim 10^{-4}$--$10^{-1}$~ms (see Fig.~\ref{fig2}B), which implies that the instantaneous diffusivity fluctuates intrinsically on the corresponding timescale.
Fluctuations of the diffusivity are negligible at the short and long timescales, where the RSD decays with $t^{-0.5}$.
The short timescale depends on the initial diffusivity $D(t = 0)$, while the long timescale relates to the relaxation time of the effective diffusivity.
In a fluctuating diffusivity model where diffusivity dichotomously fluctuates between fast and slow states~\cite{MiyaguchiAkimotoYamamoto2016, AkimotoYamamoto2016}, the magnitude of the RSD depends on the difference in diffusion coefficients between the two states and the mean residence time of states.
The magnitudes of the RSD of models 4 and 5 are higher than that of model 2.

To clarify the origin of the difference in RSDs, Fig.~\ref{fig2}C shows the PDFs of the $c$ for each model.
The PDFs of models 2, 4, and 5 have two peaks and result in large diffusivity differences between ${\rm L_o}$ and ${\rm L_d}$ domains.
Figure~\ref{fig2}D shows the PDFs of the residence times of the molecules in the ${\rm L_o}$ and ${\rm L_d}$ domains for each model. 
The residence times exhibit a power-law distribution with an exponential cutoff $P(t) \propto t^{-\beta} \exp(-t/\tau)$.
The power-law exponents for each model are almost the same, $\beta \approx -1.5$.
The cutoff in the residence time relates to the relaxation time in the RSD at which the crossover from the plateau to the $t^{-0.5}$ decay occurs.
Longer residence times of the molecule in each domain translate into longer relaxation times of the RSD.
Note that the first passage time (FPT) distribution of one-dimensional Brownian motion, starting from the origin at 0 and passing a certain point $x$, is given by the distribution $P_x(t) = |x| \exp(-x^2/4Dt) / \sqrt{4 D \pi t^3} $, that is proportional to $t^{-1.5}$ ($t \rightarrow \infty $)~\cite{KaratzasKaratzasShreveShreve1991}, where $D$ is the diffusion coefficient.
When considering a finite-sized domain, the distribution $t^{-1.5}$ has an exponential cutoff depending on the two-dimensional domain size.
The general shape of the FPT distribution is similar for many scenarios~\cite{GodecMetzler2016}.

We confirmed that slow variation of the concentration field affects little on the fluctuation of the diffusivity (see Figs.~S1 and S2).
Since the time scale of the varying field is much longer than the simulation times, the domain boundaries change slightly in equilibrium states.
In systems where the field varies faster than the time scale that particles move through the regions, a time-varying field may have a significant effect on the degree of the fluctuating diffusivity.
In addition, we note that RSDs do not depend on the field patterns (see Fig.~S3).

\begin{figure*}[tb]
\centering
\includegraphics[width = 170 mm,bb= 0 0 1276 937]{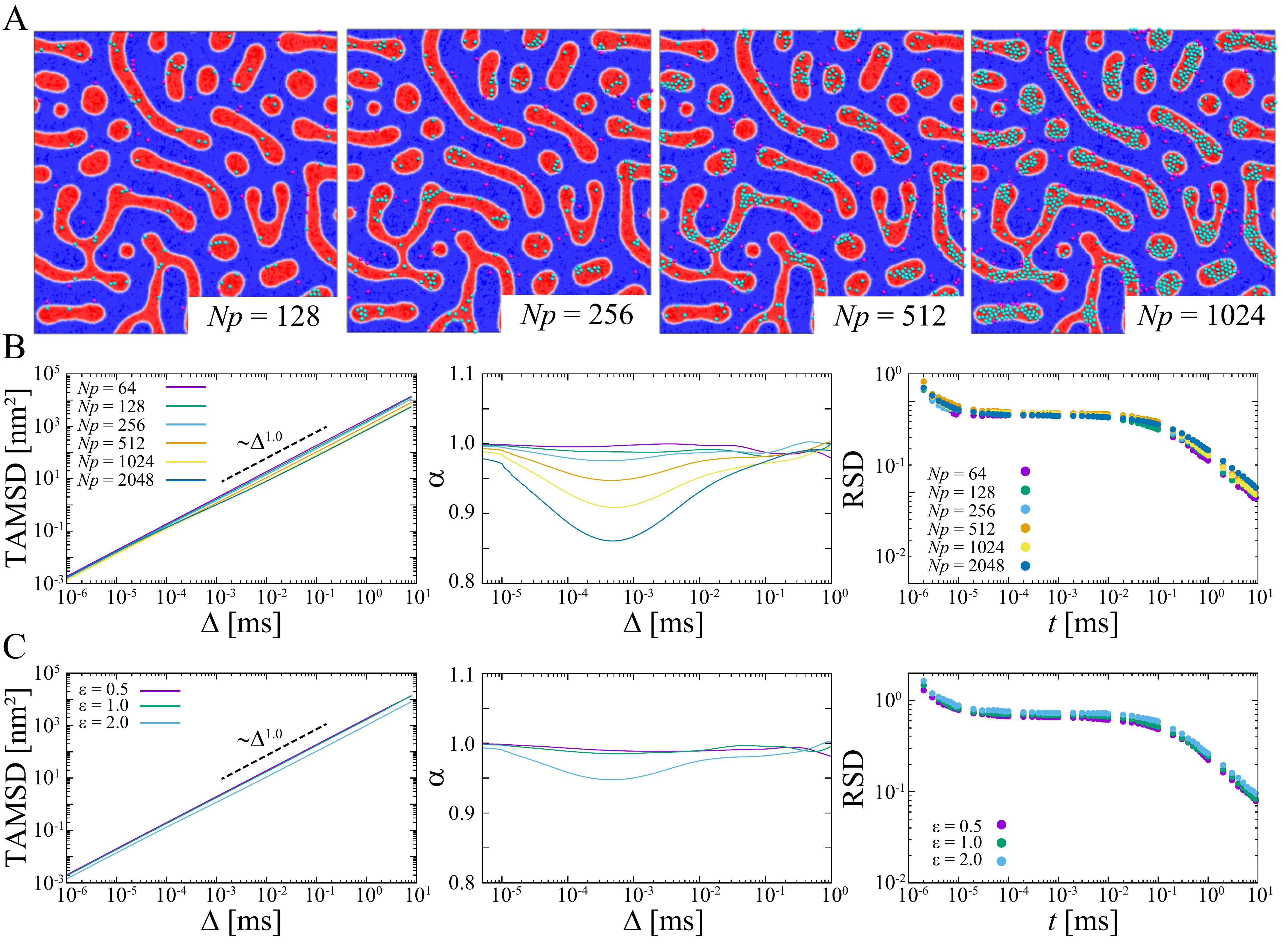}
\caption{Clustering effect of an ensemble of protein molecules on the fluctuations of the diffusivity in a heterogeneous membrane (Model5).
The number of molecules in the field $N_p$ and the interaction strength between molecules $\epsilon$ were changed.
(A)~Snapshots with different number of molecules at $N_p = 128$, 256, 512, 1024 with $\epsilon = 2.0$.
Red and blue colored regions represent ${\rm L_o}$ and ${\rm L_d}$ domains, respectively.
Molecules in ${\rm L_o}$ and ${\rm L_d}$ domains are colored cyan and magenta, respectively.
Molecules are shown with a size of $2^{1/6}\sigma$.
(B)~Ensemble averaged TAMSDs (left), time evolution of the power-law exponent $\alpha$ of the ensemble-averaged TAMSD (middle), RSD (right) compared for $N_p = 64$,128,256,512,1024,2048 with $\epsilon = 2.0$.
(C)~TAMSD, $\alpha$, and RSD for $\epsilon = 0.5$, 1.0, 2.0 with $N_p = 512$.}
\label{fig3}
\end{figure*}

\begin{figure*}[tb]
\centering
\includegraphics[width = 160 mm,bb= 0 0 1311 1200]{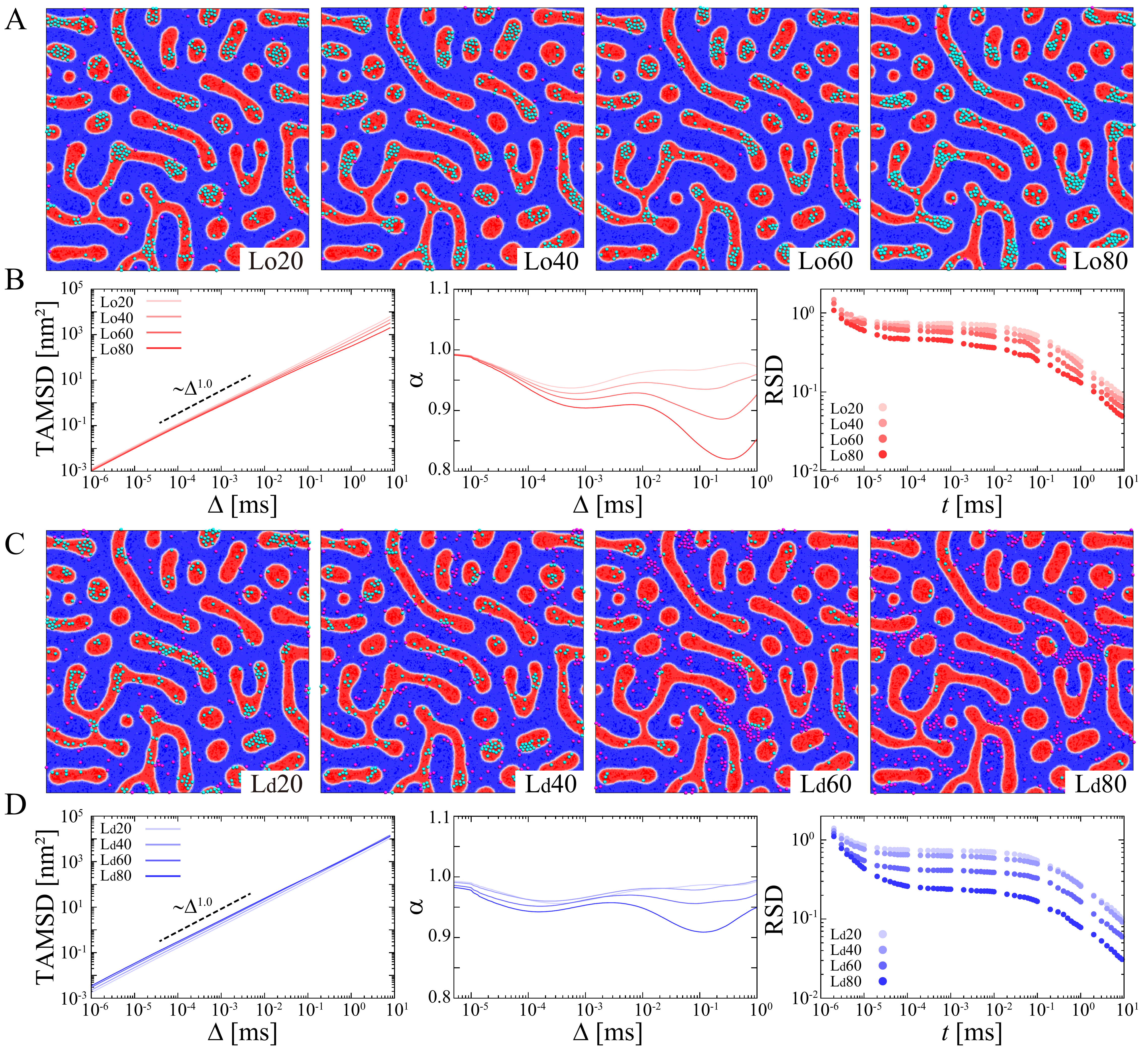}
\caption{Domain preference of protein molecules affects the diffusivity of proteins in a heterogeneous membrane (Model5).
Dependency on the degree of the domain preference (A, B)~${\rm L_o}\chi$ and (C, D)~${\rm L_d}\chi$ on diffusive dynamics of molecules.
(A, C)~Snapshots at 1~ms, (B, D)~ensemble-averaged TAMSDs, time evolution of the power-law exponent $\alpha$ of the ensemble-averaged TAMSD, and RSDs of TAMSDs are shown for each preference $\chi$.
In the snapshots, red and blue colored regions represent ${\rm L_o}$ and ${\rm L_d}$ domains, respectively.
Molecules in ${\rm L_o}$ and ${\rm L_d}$ domains are colored cyan and magenta, respectively.
Simulations were performed with $N_p = 512$ and $\epsilon = 2.0$.}
\label{fig4}
\end{figure*}

\subsection*{Clustering effect of molecules on the fluctuating diffusivity in heterogeneous membranes}
Cell membranes are crowded with a variety of proteins occupying $30$--$50~\%$ of the membrane area~\cite{GuigasWeiss2016}.
In previous studies, a concentration dependency of protein subdiffusion, $\langle  \overline{\delta {\bm r}^2(\Delta;t)} \rangle \propto \Delta ^\alpha$ with $\alpha < 1$, was observed in biological membranes~\cite{JavanainenHammarenMonticelliJeonMiettinenMartinez-SearaMetzlerVattulainen2013, JeonJavanainenMartinez-SearaMetzlerVattulainen2016}.
Switching off the protein--protein interactions changes the subdiffusive behavior ($\alpha = 0.84$) to normal diffusion ($\alpha = 1.0$)~\cite{ManzoTorreno-PinaMassignanLapeyreLewensteinGarcia2015}, dynamical correlations in the motions due to frequent molecular collisions may enhance subdiffusive motion~\cite{GuigasWeiss2016}.
In any finite system, the subdiffusive regime will ultimately cross over beyond some correlation time, see, e.g. \cite{MolinaGarciaSandevSafdariPagniniChechkinMetzler2018}.

To explore the effect of membrane crowding, we evaluate the diffusivity of molecules in molecular crowded systems with $N_p = 64$, 128, 256, 512, 1024, and 2048 particles corresponding to an area occupancy of 1.4, 3.5, 7.8, 16.6, 34.0, and 59.1~$\%$ of the ${\rm L_o}$ domains, respectively.
In the following, we mainly focus on model5 (results for other models are shown in Fig.~S4).
In this membrane state, the separation of ${\rm L_o}$ and ${\rm L_d}$ domains is most pronounced and thus best accessible in experiments.
Figure~\ref{fig3}A shows the aggregation of molecules with different area occupancy (see Movie~S1).
Even in the absence of molecular field preference, we find that as $N_p$ increases, molecules tend to aggregate in the ${\rm L_o}$ domain, where the diffusion coefficient of molecules is smaller than in the ${\rm L_d}$ domain.
This aggregation affects the diffusive behavior of molecules.
Ensemble-averaged TAMSDs become smaller and exhibit subdiffusion when the area occupancy increases (see Fig.~\ref{fig3}B).
The power-law exponent of the TAMSD decreases from $\alpha = 1.0$ to $0.85$, depending on the molecular concentration, up to a time scale of $\sim 0.1$~ms).
This trend is similar to that of MD simulations reporting transient subdiffusion of proteins in a molecular concentration-dependent manner~\cite{JavanainenHammarenMonticelliJeonMiettinenMartinez-SearaMetzlerVattulainen2013, JeonJavanainenMartinez-SearaMetzlerVattulainen2016}.
Coarse-grained MD simulation for 0.1~ms~\cite{JeonJavanainenMartinez-SearaMetzlerVattulainen2016} showed that subdiffusive motion of proteins in a noncrowded membrane changes to Brownian motion at $\Delta > 10$~ns attributed to the viscoelasticity of lipids~\cite{FlennerDasRheinstadterKosztin2009, AkimotoYamamotoYasuokaHiranoYasui2011, JeonMonneJavanainenMetzler2012, BakalisHofingerVenturiniZerbetto2015}, while in a crowded membrane significant subdiffusive regimes $\alpha \sim 0.8$--$0.9$ extends until tens of microseconds ($> 0.01$~ms is not conclusive due to the limited simulation time).
Similar transient subdiffusion induced by molecular crowding was also observed for hard-core particles~\cite{KlettCherstvyShinSokolovMetzler2021}.

Here, we discuss the effect of the concentration and the interaction strength of the tracer molecules.
Molecular concentration has little effect on the magnitude of RSD, and the relaxation time increases slightly with increasing molecular concentration.
The molecular concentration differences have little effect on the fluctuation of the diffusivity (see Fig.~\ref{fig3}B).
Moreover, we evaluated the effect of the interaction strength $\epsilon$ between molecules (see eq.[\ref{eq:LJ}] in Method). 
When the molecular interaction becomes stronger, the magnitude of the TAMSD becomes small and $\alpha$ decreases (see Fig.~\ref{fig3}C). 
An increase in the interaction strength has little effect on the RSD.

\subsection*{Preference of the domain affects the diffusivity in heterogeneous membranes}
It is known that the differences in lipid compositions in ${\rm L_o}$ and ${\rm L_d}$ domains generate a preferential partitioning of membrane proteins in either domain.
The protein domain preference, especially of transmembrane proteins, is determined by its palmitoylation, hydrophobic length, and surface area of its transmembrane region~\cite{LorentDiaz-RohrerLinSpringGorfeLeventalLevental2017, LinGorfeLevental2018}.

In our simulations, the preference was modeled using a reflective wall at the boundary between ${\rm L_o}$ and ${\rm L_d}$ domains (see details in Methods).
We evaluate the effect of preference of the ${\rm L_o}$ domain $({\rm L_o}\chi)$ (Fig.~\ref{fig4}A) or the ${\rm L_d}$ domain $({\rm L_d}\chi)$ (Fig.~\ref{fig4}B) on the diffusive dynamics, where $\chi$ is the degree of the domain preference.
As shown in Fig.~\ref{fig4}A, molecules are localized more in the ${\rm L_o}$ domain with strong ${\rm L_o}$ domain preference.
According to an increase of $\chi$, the TAMSD decreases, and the molecules exhibit more pronounced subdiffusion with smaller anomalous exponents $ \alpha =0.8$--$1.0$.
In the case of ${\rm L_d}$ domain preference, molecules are localized more in the ${\rm L_d}$ domain, and the TAMSD increases with higher $\chi$ (Fig.~\ref{fig4}B).
Molecules exhibit subdiffusion with anomalous exponents $ \alpha =0.9$--$1.0$.
Note that the crossover of $\alpha < 1$ to normal diffusion is not observed for larger ${\rm L_o}\chi$ in the studied time window (Fig.~\ref{fig4}A).
This means that the caging effect of molecules in narrow regions strongly influences anomalous diffusion, significantly more than the crowding effect with high concentrations (time scale of $\sim 0.1$~ms in Fig.~\ref{fig3}).

The magnitude of the RSDs for both ${\rm L_o}\chi$ and ${\rm L_d}\chi$ becomes smaller upon increase of $\chi$ (Fig.~\ref{fig4}).
This is thought to be due to the fact that high $\chi$ increases the confinement of molecules to a preferable domain, which leads to a decrease in the fluctuation of diffusivity.
Moreover, the area of the ${\rm L_d}$ domain is larger than that of the ${\rm L_o}$ domain in model5.
The residence time of the molecule increases with growing domain area, and the diffusivity of the molecule remains the same, resulting in a smaller RSD value.
Note that in model2 and model4 membranes, where the areas of ${\rm L_d}$ and ${\rm L_o}$ domains are the same (see Fig.~\ref{fig2}C), the change in RSD when domain preference is changed is almost the same for ${\rm L_o}\chi$ and ${\rm L_d}\chi$ (see Fig.~S4).

\begin{figure*}[tb]
\centering
\includegraphics[width = 130 mm,bb= 0 0 1099 750]{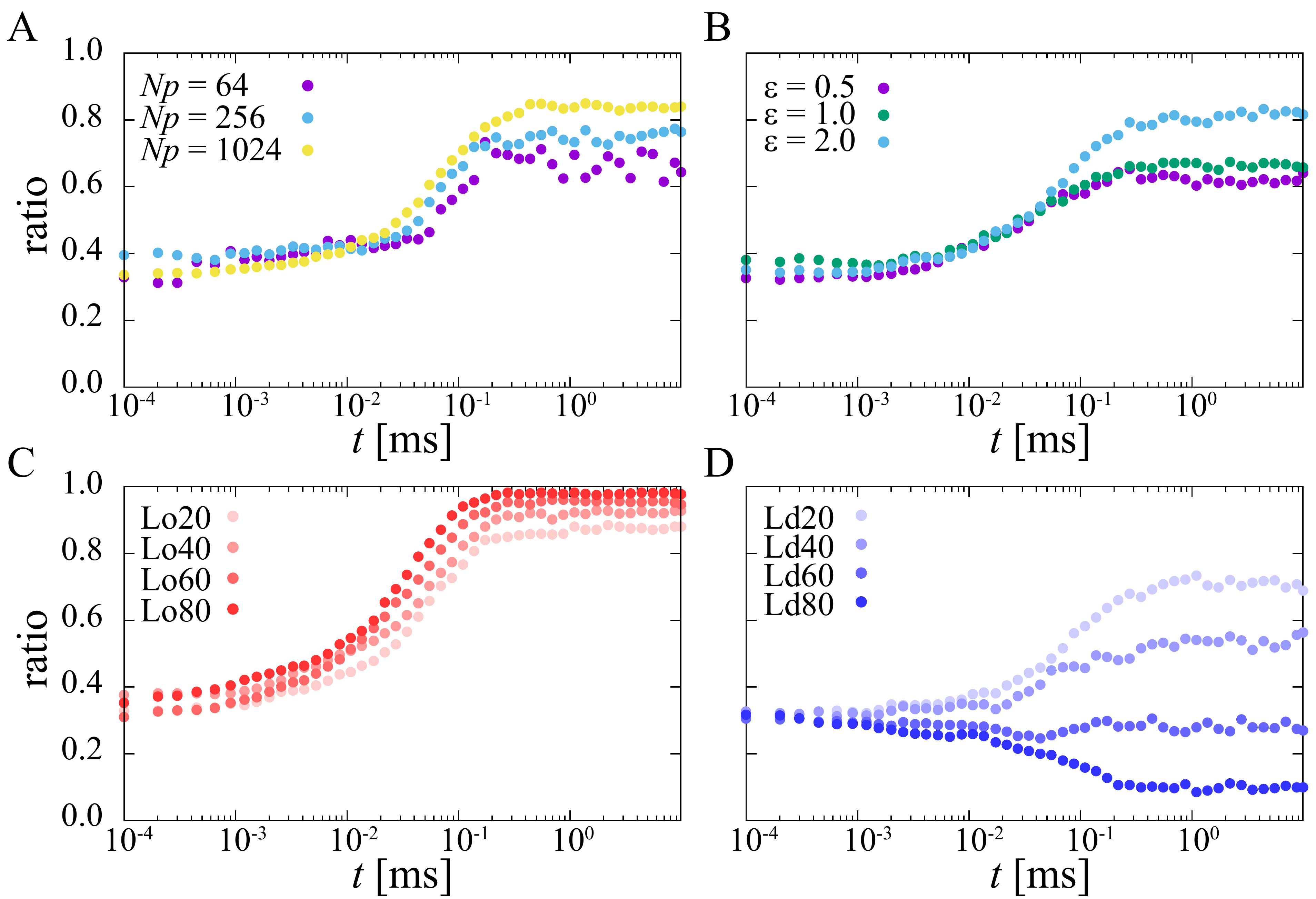}
\caption{Time variation of ratios of protein molecules confined in the ${\rm L_o}$ domain in a heterogeneous membrane (Model5).
At $t=0$, molecules were randomly distributed.
Three parameters were examined; (A)~molecular concentration ($N_p = 64, 256, 1024$) with $\epsilon = 2.0$ and ${\rm L_o}0({\rm L_d}0)$, (B)~interaction strength between molecules ($\epsilon = 0.5, 1.0, 2.0$) with $N_p=512$ and ${\rm L_o}0({\rm L_d}0)$, and (C)~molecular domain preference of ${\rm L_o}20$ -- ${\rm L_o}80$ and (D)~${\rm L_d}20$--${\rm L_d}80$ with $\epsilon = 2.0$ and $N_p=512$.}
\label{fig5}
\end{figure*}

\subsection*{Confinement of molecules to one domain due to membrane heterogeneity}
A nanoscale domain in membranes increases local molecular concentrations and molecular collisions, which are relevant to biological reactions.
To see this, the distribution of molecules in the heterogeneous membrane was analyzed.
Figure~\ref{fig5} shows ratios of molecules confined in the ${\rm L_o}$ domain examined for each parameter, such as $N_p$, $\epsilon$, and domain preference.
Randomly distributed particles at the initial time ($t=0$) diffuse and start to enrich in the ${\rm L_o}$ domain times of $t=0.01$ to $0.1$~ms.
The confinement ratio changes like a sigmoidal curve and reaches a plateau (equilibrium) after $0.1$~ms (see Fig.~\ref{fig5}AB).
Although there is no preferential domain for molecules (${\rm L_o}0$ and ${\rm L_d}0$), molecules are more likely to stay in the ${\rm L_o}$ domain, where the diffusion coefficient is lower than in the other domain, and aggregate with surrounding molecules there.
An increase in molecular concentration enhances the speed of the ratio increase and the equilibrated ratio because of high encounter rates at high concentrations (see Fig.~\ref{fig5}A).
The confinement into the ${\rm L_o}$ domain is also enhanced by the interaction strength $\epsilon$ between molecules (see Fig.~\ref{fig5}B).
The strength of $\epsilon$ does not affect the speed of the ratio increase but increases the ratio at the plateau ($t > 1$~ms) as a high interaction strength stabilizes the cluster of molecules.

We now examine the effect of domain preference, ${\rm L_o}\chi$ or ${\rm L_d}\chi$.
According to an increase in the degree of preference $\chi$, once the molecules enter the preferable domain, molecules cannot easily exit from the domain.
Figure~\ref{fig5}C shows that an increase of $\chi$ of ${\rm L_o}$ domain increases both the equilibrated ratio and the speed of the ratio.
While an increase of $\chi$ of ${\rm L_d}$ decrease the ratio of molecules in the ${\rm L_o}$ domain with a crossover around $\chi \sim 60$ (see Fig.~\ref{fig5}D).
The preferential distribution of molecules to the domain of low diffusivity is inverted by the affinity strength between molecules and the domain of high diffusivity.

\begin{figure*}[tb]
\centering
\includegraphics[width = 160 mm,bb= 0 0 547 376]{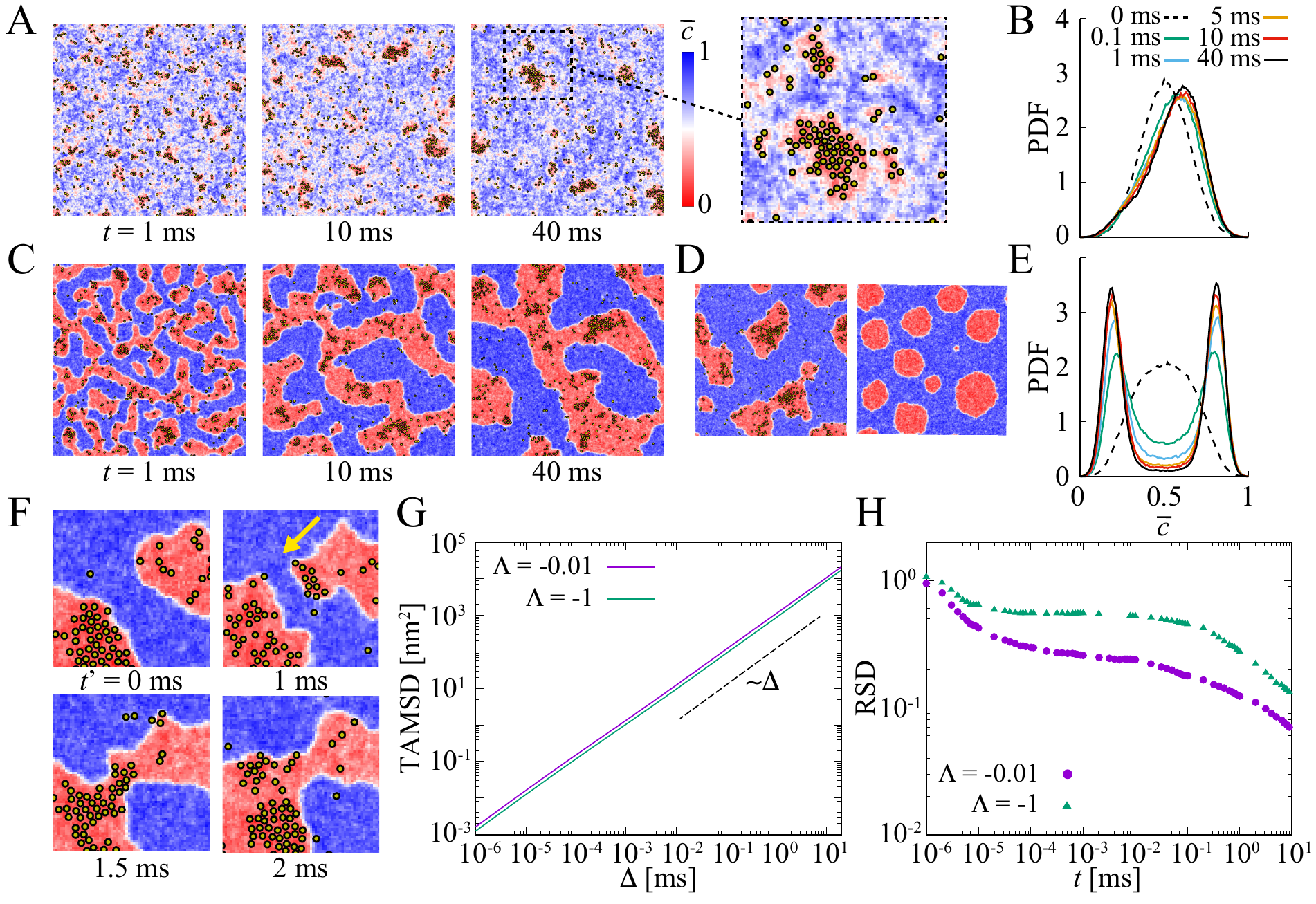}
\caption{Field modification by diffusing protein molecules: A scenario where the protein--lipid interaction induces the nanoscale phase separation and creates functional domains in the membrane.
Snapshots of the phase-separated field after 1, 10, and 40 ms have elapsed from the initially homogeneous mixed state at 0~ms, and distribution of the normalized $\overline{c}$ field: (A)(B)~a state where macroscopic phase separation does not occur spontaneously without molecules ($\Lambda = -0.01$), (C)(E)~a state where macroscopic phase separation occurs spontaneously without molecule ($\Lambda = -1$).
The initial $c$ field was set to be homogeneous with a Gaussian distribution with mean 0 and variance 1.
Diffusing molecules are represented with yellow colored circles.
(D)~Comparison of the $\overline{c}$ field after 40 ms, considering the influence of diffusing molecules on the field (left) and not considering the influence (right).
The initial homogeneous $c$ field was given with a Gaussian distribution of mean 0.3 and variance 1.
(F)~Expansion and fusion of domain regions caused by diffusing molecules.
(G)~Ensemble averaged TAMSDs for measurement time $t = 30$~ms and (H)~RSDs of TAMSDs of 512 diffusing molecules in the fields shown in (A) and (C).}
\label{fig6}
\end{figure*}

\section*{Discussion}
Visualizing small and highly dynamic domains in cell membranes can be challenging due to the reduction in binary contrast of the heterogeneity when averaging over time.
Despite prolonged debate~\cite{LeventalLeventalHeberle2020, ShawGhoshVeatch2021}, recent studies have provided compelling evidence for the coexistence of ${\rm L_o}$ and ${\rm L_d}$ domains in living cells~\cite{ToulmayPrinz2013, StoneShelbyNunezWisserVeatch2017, KinoshitaSuzukiMatsumoriTakadaAnoMorigakiAbeMakinoKobayashiHirosawaFujiwaraKusumiMurata2017, Koyama-HondaFujiwaraKasaiSuzukiKajikawaTsuboiTsunoyamaKusumi2020, UrbancicSchiffelersJenkinsGongSantosSchneiderOBrienBallVuongAshmanSezginEggeling2021, BagWagenknechtWiesnerLeeShiHolowkaBaird2021, ShelbyCastelloSerranoWisserLeventalVeatch2023}.
The mobility and aggregation of membrane proteins in cell membranes are closely linked to the local lipid order, i.e. phase separation is an organizing principle for membrane protein partitioning~\cite{UrbancicSchiffelersJenkinsGongSantosSchneiderOBrienBallVuongAshmanSezginEggeling2021}.
Recent experiments using a broad range of fluorescent probes with various membrane anchors, which have different lipid domain preferences, show the existence of segregated domains selectively partitioning membrane proteins according to their affinity for the ${\rm L_o}$ or ${\rm L_d}$ domain~\cite{ShelbyCastelloSerranoWisserLeventalVeatch2023}.
Some membrane proteins are enriched in the nanoscale region surrounding the clustered receptors.
Although these studies on multi-component systems provide us with insights into macroscopic biological regulation through heterogeneity in membranes, the underlying intricate mechanisms regulating the molecular dynamics in such complex systems have not been fully dissected.
A theoretical understanding is crucial to gain insights into the precise factors that control molecular diffusion and localization within inhomogeneous fields.
Such a quantitative model is also indispensable for data analysis in membrane systems with advanced assimilation methods based on prior training~\cite{Munoz-GilVolpeGarcia-MarchAghionArgunHongBlandBoConejeroFirbasOrtsGentiliHuangJeonKabbechKimKowalekKrapfLoch-OlszewskaLomholtMassonMeyerParkRequenaSmalSongSzwabinskiThapaVerdierVolpeWideraLewensteinMetzlerManzo2021, SecklerMetzler2022}.

In this study, we have used a well-defined in silico setup simulating molecules with fluctuating diffusivity in phase-separated fields with ${\rm L_o}$ (low diffusivity) and ${\rm L_d}$ (high diffusivity) domains.
This coarse-grained approach allows us to disentangle the various effects conspiring in the complex observed motion.
We showed that the degree of fluctuating diffusivity depends on the magnitude of the difference in molecular diffusivity between domains and the residence time in domains.
Our results suggest that molecular localization within ${\rm L_o}$ (low diffusivity) domains spontaneously occurs in heterogeneous membranes even when there is no domain preference, and subdiffusive behavior is observed due to molecular collision via molecular crowding in ${\rm L_o}$ domains.
Domain preference extends the time scale of the subdiffusive regime via molecular confinement into the preferential domains.
The effect of heterogeneity on protein partitioning was also quantitatively evaluated.
We demonstrated that the localization of molecules is determined by the difference in molecular diffusivity between domains, molecular preference of domain, and molecular concentration.

The aforementioned results were obtained under the condition of a fixed field variation to dissect the effect of pre-existing raft domains on the molecular behavior.
Membrane proteins possess the ability to modify their surrounding lipid environment, leading to the formation of functional protein--lipid complexes.
Our approach could also be applicable in scenarios where protein--lipid interactions promote or alter the functional domains within the membrane.
To model such effects, we conducted simulations under two distinct conditions: one representing a scenario where macroscopic phase separation does not occur spontaneously without proteins (near the miscibility critical point, $\Lambda = -0.01$) (Fig.~\ref{fig6}A), and another representing a state where macroscopic phase separation occurs spontaneously without proteins (under the miscibility critical point $\Lambda = -1$) (Fig.~\ref{fig6}C).
Figure~\ref{fig6}A shows snapshots of the phase-separated field driven by diffusing proteins.
Even at a condition where the macroscopic phase separation does not occur spontaneously, from a uniformly distributed state of the field and proteins, the formation of small-scale clusters of proteins leads to local ${\rm L_o}$ domains.
At this simulation condition, the normalized $\overline{c}$ has clear one distinct peak around $\sim 0.6$ (${\rm L_d}$) and rudder point at $\overline{c}<0.5$ (${\rm L_o}$) (Fig.~\ref{fig6}B).
Note that the contrast between ${\rm L_d}$ and ${\rm L_o}$ and the domain size depend on the protein--protein and protein--lipid interaction strength.

At a condition under the critical miscibility temperature, macroscopic phase separation into ${\rm L_d}$ and ${\rm L_o}$ occurs with marked contrast (Fig.~\ref{fig6}C).
The PDF of $\overline{c}$ has two peaks around $\sim 0.8$ (${\rm L_d}$) and $\sim 0.2$ (${\rm L_o}$) (Fig.~\ref{fig6}E).
This relates to a scenario of the recruitment of additional proteins to the existing functional domains and their subsequent alteration of the domain configuration and function.
Interestingly, the presence of diffusing molecules causes the distorted domain configuration, while the absence of molecules results in the formation of spherical domains (Fig.~\ref{fig6}D).
In addition, diffusing molecules cause the extension and fusion of the domains (Fig.~\ref{fig6}F).
Such recruitment can alter the thermodynamic stability of the membrane domains without a change in the lipid composition.

MSD (Fig.~\ref{fig6}G) and RSD (Fig.~\ref{fig6}H) are consistent with the previous results (fixed field variation) that magnitude of the RSD depends on the difference in diffusion coefficients between the ${\rm L_o}$ and ${\rm L_d}$ domains.

In realistic biological membranes, lipid compositions vary significantly for different cell types.
External ions and biomolecules further moderate the heterogeneity in the signaling and trafficking processes.
These factors regulate the heterogeneity of the phase-separated membrane and formation of functional multi-protein units in membranes~\cite{SakaHonigmannEggelingHellLangRizzoli2014}.
Interaction with the underlying actin cytoskeleton also regulates condensation of the phase along the actin filament by pinning elements to a preferred phase~\cite{ParmrydArumugamBassereau2015}.
In suitable conditions, the field-dependent diffusive behavior of molecules is expected to regulate the search time of partners and reaction rates~\cite{LanoiseleeMoutalGrebenkov2018}.
Protein condensation on the phase-separated membrane surfaces is a key role in downstream signaling~\cite{ChenOhBiswasZaidelBarGroves2021, ChungHuangCarboneNockaParikhValeGroves2021}.
A quantitative and qualitative elucidation of the nature of molecular behaviors in heterogeneous media is critical to understanding cellular behavior.

Our approach presented here is quite general and can be applied to fundamental questions on molecular dynamics in a variety of heterogeneous media in biology, soft matter, solid-state physics, etc.
Our model could also be extended to more realistic biological membrane models including, e.g. dynamic modulation of protein domain preferences via phosphorylation by interaction with regulatory proteins~\cite{LeventalLyman2023}, protein remodeling through conformational changes, complex inhomogeneous interaction between molecules and clusters, the partitioning by an actin filament mesh~\cite{KalayFujiwaraKusumi2012, SadeghHigginsMannionTamkunKrapf2017}, alternation of membrane composition in signaling events, and the partitioning regulation of membrane signaling~\cite{YouMarquez-LagoRichterWilmesMoragaGarciaLeierPiehler2016}.
Moreover, the parameters of this mesoscale simulation can be determined bottom-up from MD simulations, allowing comparison of mesoscopic molecular behavior at the intersection of simulations and experimental spatiotemporal scales.
Concurrently, advanced single particle tracking studies provide massive new data on protein dynamics in membranes that can be scrutinized by our approach and advanced methods for data analysis~\cite{Munoz-GilVolpeGarcia-MarchAghionArgunHongBlandBoConejeroFirbasOrtsGentiliHuangJeonKabbechKimKowalekKrapfLoch-OlszewskaLomholtMassonMeyerParkRequenaSmalSongSzwabinskiThapaVerdierVolpeWideraLewensteinMetzlerManzo2021, SecklerMetzler2022}.
This could open a new direction to delineate the role of heterogeneity in the membranes with more complex multicomponent systems in a more physiological setting~\cite{GoswamiGowrishankarBilgramiGhoshRaghupathyChaddaVishwakarmaRaoMayor2008, GowrishankarGhoshSahaRumamolMayorRao2012}.

\section*{Methods}
\subsection*{Simulation models}
We used five models of phase separation in cell membranes as described in Ref.~\cite{FanSammalkorpiHaataja2010}.
The specific choices for the parameters represent lipid raft formation (model1) by thermal fluctuations near the critical temperature, (model2) by pinning of the interfacial composition of immobile membrane proteins~\cite{YethirajWeisshaar2007, LaradjiGuoGrantZuckermann1992}, (model3) in miscible or (model4) immiscible lipid systems, and (model5) by exchange with lipid reservoirs~\cite{GomezSaguesReigada2008, Foret2005, FanSammalkorpiHaataja2010a}.
These five models are expressed using a Cahn-Hilliard equation~\cite{FanSammalkorpiHaataja2010, FanSammalkorpiHaataja2010b, BerryBrangwynneHaataja2018} for the order parameter field $c({\bm r},t)$,
\begin{equation}
    \frac{\partial c({\bm r},t)}{\partial t} = - \frac{1}{\tau_r} (c - c_r)
    + M \nabla^2 \frac{\delta F}{\delta c} + \eta ({\bm r},t).
    \label{eq:chan}
\end{equation}
The first term on the right-hand side is the term for the lipid reservoir in model5, where \(\tau_r\) is a parameter representing the relaxation time due to coupling to the lipid reservoir, and $c_r$ is the average compositions imposed by the lipid reservoir.
The second term on the right-hand side is the modified Ginzburg-Landau free energy term in the usual Cahn-Hilliard equation, where $M$ is the mobility and $F$ is the free energy,
\begin{equation}
    F = \int \left\{\frac{W^2}{2} [1 - \alpha \rho ({\bm r})] (\nabla c)^2
    + \frac{\Lambda c^2}{2} + \frac{c^4}{4}\right\} d{\bm r},
    \label{eq:free_energy}
\end{equation}
where the parameter $\Lambda > 0$ $(\Lambda < 0)$ is a relative temperature to the mean-field critical temperature $T>T_c$ $(T<T_c)$.
$W$ is a parameter to control the line tension between the raft and nonraft phases.
$\alpha$ is a parameter that explain the local reduction in the line tension due to immobile membrane proteins.
The local concentration $\rho ({\bm r})$ of $N$ immobile membrane proteins in model2 can be expressed as,
\begin{equation}
    \rho ({\bm r}) = \pi \sigma^{-2}_{\rm IMP} \sum^N_i {\rm exp}
    \left(- \frac{|{\bm r} - {\bm r}_i|^2}{2 \sigma^2_{\rm IMP}} \right) .
\end{equation}
$\eta ({\bm r},t)$ in eq.~\ref{eq:chan} denotes a Gaussian noise term~\cite{FanSammalkorpiHaataja2010b}, 
\begin{equation}
    \eta ({\bm r},t) = {\cal F}^{-1} \left[
    \frac{(H \sqrt{\Delta t} / \Delta x) l |{\bm q}|}{\sqrt{1 + {\bm q}^2 l^2}}
    \times \hat{\xi}({\bm q},t) \right],
\end{equation}
where $H$ can be related to either the temperature of the system or the rate at which lipids are removed and added to the leaflet due to vesicular and nonvesicular lipid trafficking events, $l$ denotes the recycling length over which spatial redistribution of lipids takes place~\cite{FanSammalkorpiHaataja2008}, and \(\hat{\xi} ({\bm q},t)\) is the Fourier transform of the white Gaussian noise with mean 0 and variance 1.
Here, we used $c_r = 0.3$, $M=1$, $\sigma_{\rm IMP} = 1/\sqrt{2}$, $W=1$, and the values of each parameter in each membrane model are shown in Table \ref{table:miko}~\cite{FanSammalkorpiHaataja2010}.
The ordered ($c<0$) and disordered ($c>0$) phases denote the raft (${\rm L_o}$) and non-raft (${\rm L_d}$) domains.

\begin{table*}[tb]
    \caption{Values of the parameters in each model~\cite{FanSammalkorpiHaataja2010}.}
    \label{table:miko}
    \centering
    \begin{tabular}{|c|c|c|c|c|c|c|c|}
        \hline
        model & \(t_r\)    & \(\Lambda\) & \(l\) & \(H\)  & \(\alpha\) & \(N\) & number of steps \\
        \hline
        1      & \(\infty\) & -0.001      & 0.1   & 0.0283 & 0          & 0     & 24000      \\
        \hline
        2      & \(\infty\) & -1          & 1     & 0.85   & \(1/\pi\)  & 1500  & 1500000    \\
        \hline
        3      & \(\infty\) & 10          & 1280  & 0.85   & 0          & 0     & 180        \\
        \hline
        4      & \(\infty\) & -1          & 1280  & 0.85   & 0          & 0     & 24000      \\
        \hline
        5      & 500        & -1          & 0.1   & 2.12   & 0          & 0     & 720000     \\
        \hline
    \end{tabular}
\end{table*}

The system was simulated using the phase-field method under periodic boundary conditions with a grid point size of $256 \times 256$ (256~nm $\times$ 256~nm in physical dimensions).
The lattice point width was set to $\Delta x = \Delta y = 1$ (1~nm for physical quantities).
The time step was set to $dt = 0.005$ for dimensionless numbers, which corresponds to $10^{-5}$~s for physical quantities.
The number of simulation steps for each model is shown in Table~\ref{table:miko}.
The initial $c$ field was set to be homogeneous with a Gaussian distribution with mean 0 and variance 1.

\subsection*{Single particle system}
The diffusive particles in each membrane model are modeled by the Langevin equation \ref{eq:langevin} with fluctuating diffusivity.
The diffusivity of the particle, $D({\bm r}(t),t) = (c_b + \overline{c({\bm r}(t),t)}) D_0$, fluctuates depending on the normalized order parameter field $\overline{c({\bm r}(t),t)}$ ($0 < \overline{c} <1$).
We used $\overline{c({\bm r}(t),t)}$ in equilibrium after running simulations for each number of steps in  Table~\ref{table:miko}.
The single-particle simulations were performed 100 times with different initial coordinates of the particles for the same phase-separated field.
The parameters $c_b = 1$ and $D_0 = 1$ (1~$\mu$m$^2$/s) were used in each model.
Simulations were performed for $10^7$ steps (10~ms) with $dt = 10^{-3}$ (1~ns), and the trajectories of the particles were analyzed after $10^6$ steps (1~ms) of reaching equilibrium.

\subsection*{Multi particle system}
For multi particle interactions, we performed simulations including particle-particle interactions,
\begin{equation}
    \frac{d{\bm r}(t)}{dt} = - \frac{D({\bm r}(t),t)}{k_B T} \frac{ d U(l)}{d l} + \sqrt{2D({\bm r}(t),t)} \omega(t) ,
     \label{eq:lang-multi}
\end{equation}
where, $k_B T = 1$, and Lennard-Jones potential was used,
\begin{equation}
    U(l) =4 \epsilon  \left\{ \left( \frac{\sigma}{l} \right)^{12} - 
    \left( \frac{\sigma}{l} \right)^6 \right\}
    \label{eq:LJ}
\end{equation}
where $l$ was the distance between two interacting particles, size of the particle $\sigma$ was $3.0$.
The depth of the potential well $\epsilon$ was set as 0.5, 1.0, 2.0.
The number of particles in the system $N_p$ was set to $N_p = 64, 128, 256, 512, 1024, 2048$ to compare the effect of particle concentration on the diffusivity.
For multiple particle systems, we used $c_b=0.1$ and $D_0 = 1$ in $D({\bm r}(t),t) = (c_b + \overline{c({\bm r}(t),t)}) D_0$.

We used cdview (https://polymer.apphy.u-fukui.ac.jp/ \textasciitilde koishi/cdview.php) for visualization of the simulations.

\subsection*{Nanoscale phase separation model modified by diffusive particles}
For a scenario where protein--lipid interaction induces nanoscale phase separation and creates functional domains in the membrane, we conducted a simulation in which diffusing particles change the field.
For phase separation, we considered the following expression,
\begin{equation}
     \frac{\partial c({\bm r},t)}{\partial t} = \nabla  \left\{ M \nabla \left[ - \nabla^2 c + \Lambda c + c^3 +  g({\bm r},t) \right] \right\} + \eta ({\bm r},t),
    \label{eq:pf-vary}
\end{equation}
where $M=1$, $g({\bm r},t)$ is a term of short-ranged protein--lipid interaction,
\begin{equation}
g({\bm r},t) =
\begin{cases}
\alpha_g&\text{($r \leq \sigma$)}\\
\alpha_g \exp \left( -\frac{r-\sigma}{r_g} \right)&\text{($r > \sigma$)}
\end{cases}.
\end{equation}
$g({\bm r},t)$ was considered at the positions of the diffusing particles.
Here, we set the intensity of field modification by proteins as $\alpha_g = 0.5$ and its relaxation length as $r_g = 2$.

The particles were simulated using eq.~\ref{eq:lang-multi} and eq.~\ref{eq:LJ} with parameters $\sigma = 3.0$, $\epsilon = 2.0$, and $N_p = 512$.
The $c$ field was updated every 10 steps of the Langevin simulation (eq.~\ref{eq:lang-multi}).
Simulations were performed for $4\times 10^7$ steps (40~ms) with $dt = 10^{-3}$ (1~ns), and the trajectories of the particles were analyzed after $10^7$ steps (10~ms).

\subsection*{Domain preference of molecules}
To implement the domain preference of the molecule, the energy barrier between the domains was reproduced by probabilistic reflection when a particle moves from one domain to another.
We compared three patterns.
One is that the particles can move freely between the ${\rm L_o}$ and ${\rm L_d}$ domains without being reflected ($\chi = 0$).
The other two are cases where the particles exhibit ${\rm L_o}$ or ${\rm L_d}$ preferences.
${\rm L_o}\chi$ means that the molecule is reflected at ${\rm L_o}$ when moving from the ${\rm L_o}$ domain to the ${\rm L_d}$ domain at $\chi$~\% probability of reflection and not reflected when moving in the opposite direction.
${\rm L_d}\chi$ is vice versa.

\subsection*{Acknowledgments}
This work was supported by JSPS KAKENHI Grant Number 20K14432.
M.S. thanks Wellcome (grant no. 208361/Z/17/Z) for support.
R.M. thanks the German Science Foundation (DFG, grant no. ME1535/12-1) for support.


%

\end{document}